\documentclass[%
 reprint,
preprintnumbers,
 amsmath,amssymb,
 prl,
]{revtex4-2}

\usepackage{graphicx}
\usepackage{dcolumn}
\usepackage{bm}

\usepackage{amsthm}
\newtheorem{thm}{Theorem}
\usepackage{mathrsfs}
\usepackage{physics} 
\usepackage{tensor}
\usepackage{xcolor}

\newcommand{\PI}{PI\ }
\newcommand{\PInospace}{PI}

\begin{document}

\preprint{MIT-CTP/5458}

\title{The Penrose Inequality as a Constraint on the Low Energy Limit
of Quantum Gravity}

\author{\AA{}smund Folkestad}
\affiliation{
 Center for Theoretical Physics, Massachusetts Institute of Technology,
 Cambridge, MA 02139, USA \\
Kavli Institute for Theoretical Physics, University of California, Santa
Barbara, CA 93106, USA
}


\begin{abstract}
    We construct initial data violating the Anti-deSitter Penrose inequality using scalars
    with various potentials. Since a version of the Penrose inequality can be derived from
    AdS/CFT, we argue that it is a new swampland
    condition, ruling out holographic UV completion for theories that violate
    it. We produce exclusion plots on scalar couplings violating the inequality,
    and we find no violations for potentials from string theory.  In the special case
    where the dominant energy condition holds, we use GR techniques to prove the
    AdS Penrose inequality in all dimensions, assuming
    spherical, planar, or hyperbolic symmetry. However, our violations show that
    this result cannot be generically true with only the null energy condition,
    and we give an analytic sufficient condition for violation of the Penrose
    inequality, constraining couplings of scalar
    potentials. Like the Breitenlohner-Freedman bound, this gives a necessary 
    condition for the stability of AdS.
\end{abstract}

\maketitle

\textit{Introduction.}
Whether or not singularities are hidden behind event horizons is a
 longstanding open question in general relativity. In
\cite{Pen73} Penrose showed that if (1) the answer to this is
affirmative, and (2) collapsing matter settles down to Kerr, then the
existence of certain special surfaces $\sigma$ appearing in regions
of strong gravity implies a lower bound on the spacetime mass:
\begin{equation}\label{eq:vanillaPI}
\begin{aligned}
    G_N M \geq \sqrt{\frac{ \text{Area}[\sigma] }{ 16\pi }}.
\end{aligned}
\end{equation}
A proof of this inequality, named after Penrose, would amount to evidence in
favor of singularities being hidden, but the inequality has not been
proven except in special cases \cite{HuiIlm01,Bra01} (see \cite{Mar09}
for a review).

Recently, Engelhardt and Horowitz \cite{EngHor19} gave a holographic argument
for an AdS version of the Penrose inequality (PI), assuming the AdS/CFT correspondence, but not cosmic
censorship nor anything about the endpoint of gravitational collapse. This
suggests that hypothetical bulk matter allowing violation of the PI in AdS is
incompatible with the AdS/CFT dictionary, and that the PI can serve as
a new condition detecting low energy theories that cannot be UV completed in holographic
quantum gravity, meaning theories that can never arise as the low-energy limit of a holographic quantum gravity
theory valid at all energy scales. 

In this article we construct violations of the PI for various scalar
potentials, and produce exclusion plots in coupling space, delineating
regions where we know that the PI is violated.  Since the PI turns out
to constrain neutral scalars, we find that it is distinct from the weak
gravity conjecture \cite{ArkMot06}. Next, we present numerical evidence that
supersymmetry is a sufficient condition for the PI. We also present an
analytical sufficient condition on scalar couplings for a theory to
violate the PI. Similar to the Breitenlohner-Freedman bound
\cite{BreFre82, BreFre82b}, this provides a necessary
condition for the stability of AdS. Next, while our work shows that
general theories respecting the null energy
condition (NEC) violate the PI, we are able to prove the PI
in all dimensions greater than two for any theory satisfying the dominant energy condition
(DEC), assuming spherical, planar, or hyperbolic symmetry. 

We emphasize that while we in this work use the PI to constrain theories in the classical limit,
these constraints are intimately
tied to quantum gravity in the form of the AdS/CFT correspondence, which
is a nonperturbative description of string theory in AdS \cite{Mal97,GubKle98,Wit98a}. This is because Penrose's 
original argument for his inequality is invalid for general low energy theories
in AdS, since there exist theories violating cosmic censorship in AdS
\cite{HorSan16,CriSan16,CriHor17,HorSan19}. The only known way to to argue for the
truth of the PI in AdS is using the full machinery of the AdS/CFT
correspondence, and then taking its classical limit. Without reference to
AdS/CFT, we have no principle to exclude theories violating the PI, while if we demand that our theory arises as the
classical/low-energy limit of holographic quantum gravity, the PI must hold.

\textit{The Penrose Inequality in AdS/CFT.}
Consider an apparent horizon $\sigma$ in an asymptotically AdS$_{d+1}$
(AAdS)
spacetime with mass $M$, meaning that the expansion of the outwards
null geodesic congruence fired from $\sigma$ is vanishing, 
while the inwards expansion is non-positive. Assuming the holographic
dictionary, Ref.~\cite{EngHor19} derived that
\begin{equation}\label{eq:AdSPI}
\begin{aligned}
    \text{Area}[\sigma] \leq A_{\rm BH}(M),
\end{aligned}
\end{equation}
where $A_{\rm BH}(M)$ is the area of the most entropic stationary black
hole of mass $M$ in the theory. This is the AdS version of the PI that can be
derived in holography, and
by knowing the function $A_{\rm BH}(M)$, Eq.~\eqref{eq:AdSPI} can be rewritten to
give a lower bound on the mass, similar to Eq.~\eqref{eq:vanillaPI} (see
Eq.~\eqref{eq:AdSPIproven}).

The argument of \cite{EngHor19} relied on (1) the HRT entropy formula 
\cite{RyuTak06, RyuTak06-2, HubRan07, LewMal13}, (2) the existence of
the so-called
coarse grained CFT state, whose von-Neumann
entropy equals $\text{Area}[\sigma]/4G_N$ \cite{EngWal17b, EngWal18}, and (3) the
fact that there exists a gravitational path integral for the microcanonical
ensemble which has stationary black holes as saddles \cite{BroYor92, Mar18}.
The argument also makes the reasonable assumption that there is no spontaneous breaking
of time translation symmetry in the CFT microcanonical ensemble, so that the
microcanonical ensemble is dual to a stationary black hole \footnote{We thank
Don Marolf for pointing this out.}.
Finally, $\sigma$ had to satisfy two technical conditions: 
that it becomes a proper trapped surface when perturbed slightly inwards,
and that $\sigma$ is outermost minimal, 
meaning that there exists a spacelike or null hypersurface bounded by
$\sigma$ and the conformal boundary on which no other surface is
smaller (see \cite{EngWal17b, EngWal18} for precise conditions). In the special
case where $\sigma$ is an extremal surface, the first condition is not needed.

\textit{Constraining Scalar Potentials.}
Working with scalar fields and spherical symmetry in the classical
limit,
we will see that many scalar
potentials that violate the DEC violate
Eq.~\eqref{eq:AdSPI}
as well.  DEC-violating scalars are important, since they appear in known 
examples of AdS/CFT dualities after dimensional reduction of compact dimensions
\cite{HerHor03,LuPop99,LuPop99b, CveDuf99}.  A generic DEC violating scalar potential will
not even have a positive mass theorem (PMT) \cite{SchYau81,Wit81}, and
in these theories the \PI is automatically violated, but we will also find that
theories where we are unable to construct negative mass solutions, despite
extensive numerical search, will frequently violate the \PInospace.

The theories we consider have the action
\begin{equation}\label{eq:scalaraction}
\begin{aligned}
    \int \dd^{d+1}x\sqrt{-g}\left[\frac{ 1 }{ 2 }R + \frac{ d(d-1) }{ 2L^2 } - \frac{ 1 }{ 2
    }|\nabla \phi|^2 - V(\phi) \right],
\end{aligned}
\end{equation}
where $L$ is a length scale that sets the cosmological constant, and where
$V$ is a potential  satisfying $V(0)=V'(0)=0$. To look for violations of the
\PInospace, we will construct AAdS initial data on a
partial Cauchy surface $\Sigma$ bounded by $\sigma$ and the conformal
boundary, such that (1) $\sigma$ is an apparent horizon satisfying all
the technical conditions relevant for Eq.~\eqref{eq:AdSPI}, and (2) $\Sigma$
can
be embedded in a larger initial dataset on a complete hypersurface.  This is sufficient
to test if the \PI holds for $\sigma$; the full spacetime is not needed. 

What scalar potentials $V(\phi)$ should we consider in order to find violations
of the PI?
Ref.~\cite{HusSin17} proved an AdS$_{4}$ PI assuming spherical symmetry and the
DEC. Assuming the ordinary gravitational mass $M$ \cite{AshDas99} is finite, 
we prove the following generalization (conjectured to be
true in \cite{ItkOz11}):
\begin{thm}\label{thm:PI}
    Consider an asymptotically AdS$_{d+1\geq3}$ spacetime with
    spherical ($k=1$), planar ($k=0$), or hyperbolic symmetry ($k=-1$),
    satisfying the Einstein equations 
    $G_{ab} - \frac{ d(d-1) }{ 2L^2 }g_{ab} = 8\pi G_N T_{ab}$ 
    and the DEC: $T_{ab}u^a v^a \geq 0$ for all timelike $u^a, v^a$.
    If $\sigma$ is a symmetric outermost marginally trapped surface with respect
    to a connected component
    of the conformal boundary with mass $M$, then
    \begin{equation}\label{eq:AdSPIproven}
    \begin{aligned}
        \frac{ 16\pi G_N }{ (d-1)\Omega_{k} } M \geq k\left(\frac{ {\rm Area}[\sigma] }{ \Omega_{k} } \right)^{\frac{
            d-2 }{ d-1 }} + \frac{ 1 }{ L^2 }\left(\frac{ {\rm Area}[\sigma] }{ \Omega_{k} } \right)^{\frac{ d }{ d-1
        }}.
    \end{aligned}
    \end{equation}
\end{thm}
\noindent Here $\Omega_{k}$ is the volume of the $(d-1)$--dimensional unit sphere, the plane, or the
unit hyperbolic space (or a compactification thereof, in the latter two cases).
While $\Omega_{k}$ might be infinite, the ratios $\text{Area}[\sigma]/\Omega_{k}$ and
$M/\Omega_{k}$ are well defined. Furthermore, taking $k=1$ and $L\rightarrow
\infty$ we get the PI for spherically
symmetric asymptotically flat space in general dimensions. The mass is
conventionally defined so $M=0$ for pure AdS (see \cite{HolIsh05} for a
discussion definitions of mass in AdS). Let us now turn to the proof.

\textit{Proof:}
Consider an AAdS$_{d+1}$ spacetime with spherical, planar, or hyperbolic symmetry, 
and consider a null gauge with coordinates $(x^+, x^-, \Omega^i)$ and metric
\begin{equation}
\begin{aligned}
    \dd s^2 = -2 e^{-f(x^+, x^-)}\dd x^+ \dd x^- + r(x^+, x^-)^2 \dd \Omega_k^2,
\end{aligned}
\end{equation}
where $r$ is a function of $(x^+, x^-)$ and 
where $\dd \Omega_k^2$ locally is the (unit) metric on the sphere,
plane, or hyperbolic space. Define $k_{\pm}^a = (\partial_{x^{\pm}})^a \equiv
(\partial_{\pm})^a$, which has associated null expansions
$\theta_{\pm}=(d-1)r^{-1}\partial_{\pm}r$. The quantity
\begin{equation}
\begin{aligned}
    \mu(x^+, x^-) = r^{d}\left[\frac{ k }{ r^2 }-\frac{ 2\theta_{+}\theta_{-} }{
        k_+ \cdot k_- (d-1)^2} + \frac{ 1 }{ L^2 }\right],
\end{aligned}
\end{equation}
can be seen to reduce to the spacetime mass at $r=\infty$, up to
an overall factor:  $16\pi G_N M = (d-1)\Omega_{k}\mu|_{r=\infty}$. 
The null-null components of the Einstein equations (in units
with $8\pi G_N
=1$) reduce to
\begin{equation}
    \begin{aligned}\label{eq:nullEFE}
    \frac{ r  T_{\pm \pm}}{ d-1 }  &= -\partial_{\pm}f \partial_{\pm}r -
    \partial_{\pm}^2 r,\\
    \frac{ r  T_{+-}}{ d-1 } &= \partial_{+}\partial_{-}r + \frac{ d^2 - 3d + 2
        }{ (d-1)r }\left[\frac{e^{-f}}{2}\left(k + \frac{ r^2 }{ L^2 } 
    \right) + \partial_{+}r \partial_{-}r \right] \\
        &\quad + \frac{ r }{ L^2 }e^{-f}
\end{aligned}
\end{equation}
Proceeding similarly to Ref.~\cite{Hay94b}, we compute $\partial_{\pm}\mu$ 
and use Eqs.~\eqref{eq:nullEFE} to eliminate $\partial_{\pm}r, \partial_{\pm}^2r,
\partial_{+}\partial_{-}r$,
yielding
\begin{equation}
\begin{aligned}
    \partial_{\pm}\mu = \frac{ 2 e^{f} r^{d} }{ (d-1)^2
    }\left(T_{+-}\theta_{\pm} - \theta_{\mp}T_{\pm\pm} \right).
\end{aligned}
\end{equation}
The DEC implies that $T_{\pm\pm}\geq 0$ and $T_{+-}\geq 0$.
Thus, $\pm \partial_{\pm}\mu$ is positive in an untrapped region ($\theta_+ \geq 0,
\theta_{-}\leq 0$), and so there $\mu$ is monotonically non-decreasing in an outwards spacelike
direction. Evaluating $\mu$ on a marginally trapped surface that can be deformed
to infinity along a untrapped spacelike path, which exists by the assumption
that $\sigma$ is outermost marginally trapped, gives that $kr^{d-2}+r^{d}L^{-2}\leq
\mu|_{r=\infty}$. Converting $\mu|_{r=\infty}$ to mass gives
Eq.~\eqref{eq:AdSPIproven}. $\square$

Now, the above proof applies for an apparent
horizon which is outermost marginally trapped, which is not always the same 
as outermost minimal. However, at a moment of time-symmetry the two always
coincide, since in this case we have that $\theta_{\pm} = \pm
\mathcal{K}$ \cite{EngFol21b}, where $\mathcal{K}$ is the mean curvature of $\sigma$ in $\Sigma$,
and minimality means that $\mathcal{K}=0$.  Thus, to look for
violations of the PI in our setup, Theorem~\ref{thm:PI} shows that we need to consider 
theories violating the DEC, which for \eqref{eq:scalaraction} means potentials that are
negative somewhere.

As mentioned, DEC violating potentials arise in known
AdS$_{d+1}$/CFT$_{d}$ dualities after dimensional reduction,
but we can also see their relevance more directly. In AdS/CFT, bulk
scalar fields are dual to local scalar operators $O(x)$ in the CFT that transform with scaling
dimension $\Delta$ under dilatations: $O(x) \rightarrow
\lambda^{\Delta}O(\lambda x)$. It turns out that whenever $O$ is
a relevant operator (i.e. $\Delta < d$), we must have that $m^2 \equiv
\partial_{\phi}^2 V(\phi)|_{\phi =0}<0$, leading to DEC violation.
This follows from the standard expression for the scaling dimension
$\Delta$ of $O$ \cite{Wit98a}: $\Delta = d/2 +
\sqrt{(d/2)^2+m^2L^2}$ \footnote{It is possible to choose boundary
conditions so that the scaling dimension of the operator dual to $\phi$ is
$\Delta = \frac{ d }{ 2 }-\sqrt{(d/2)^2+m^2L^2}$ \cite{Wit98a}. We do not
do this here, as these boundary conditions require modifications of the
definition of the spacetime mass.}. $\Delta<d$ indeed means negative
$m^2$, which is allowed as long as 
the Breitenlohner-Freedman (BF) bound \cite{BreFre82, BreFre82b} is
satisfied: $m^2 \geq m_{\rm BF}^2 \equiv -d^2/(4L^2)$. 

\textit{Black Hole Uniqueness, Positive Mass, and Compact Dimensions.}
Before constructing initial data, a few
subtleties and known results should be addressed.
First, the reference black hole of mass $M$ appearing in
Eq.~\eqref{eq:AdSPI} is the one that dominates the microcanonical ensemble
at that mass, which is the one with the largest area \cite{Mar18}.
Thus, if there exist black holes with larger
area than AdS-Schwarzschild at a given mass, we seemingly have to construct these 
before claiming a violation.  Black hole
uniqueness is not established in AdS, so this seems like a difficult task. 
However, spherical symmetry allows significant simplification. 
In the static spherically symmetric case, Ref.~\cite{XiaZi22} recently proved that the NEC implies 
    $A_{\rm BH}(M) \leq A_{\rm AdS-Schwarzschild}(M)$,
so AdS-Schwarzschild is the only spherical black hole that can dominate the microcanonical
ensemble. Since the theories we consider here respect the NEC,
we thus know that AdS-Schwarzschild is the correct black hole to compare to in
Eq.~\eqref{eq:AdSPI}, assuming we can take the reference black hole to be
spherically symmetric. This is reasonable, and amounts to the assumption
that the CFT microcanonical ensemble on a sphere does not break rotational symmetry
spontaneously (in the bulk this is the fact that introducing spin at fixed energy tends to reduce the
area, as can be seen from Kerr-AdS \cite{Car68} and other known spinning black hole
solutions \cite{Gub98,CveGub99,AltKub14}).

Second, it has been proven that the PMT holds even in certain theories
violating the DEC. The prime example is in classical supergravity (SUGRA) theories
\cite{BreFre82, GibHul83,BreFre82b}, but in Einstein-scalar theory
more general results are known. It was proved in \cite{Bou84, Tow84}
that the PMT holds if the scalar potential $V(\phi)$ can be written as
\begin{equation}\label{eq:superPot}
\begin{aligned}
    V(\phi) = \frac{ d(d-1) }{ 2L^2 } + (d-1)W'(\phi)^2 - d W(\phi)^2
\end{aligned}
\end{equation}
for some real function $W(\phi)$ defined for all $\phi \in
\mathbb{R}$ and satisfying $W'(0)=0$ (provided we only turn on the scalar mode with
fastest falloff \cite{AmsMar06,AmsHer07,FauHor07}, which is what we do here). 
If we considered a supersymmetric theory, $W$ would be the so-called superpotential,
but supersymmetry is not required, and $W$ can be any function satisfying the
above properties. Nevertheless, we keep referring to $W$ as a superpotential. It is not
known whether the existence of $W$ is a necessary condition for the
existence of a PMT; the proofs of \cite{Bou84, Tow84} only show that
it is sufficient.

Third, suppose that an AAdS$_{d+1}$ solution is a dimensional reduction of a
higher dimensional solution with some number of compact dimensions. If the
higher dimensional solution is a warped product rather than a product metric
between AAdS$_{d+1}$ and the compact space, 
then it is not a priori
obvious that a violation of the lower dimensional PI implies a
violation of the higher dimensional one. For theories stemming from
higher dimensions, it could in principle be that
the PI only is valid with all dimensions included, but our numerical findings argue against
this, since potentials from known AdS/CFT dualities seem to respect the lower dimensional
PI, as we will see 
\footnote{Furthermore, \cite{JonMar16} found compelling evidence
that $\text{Area}[\sigma]/G_N$ is the same whether is computed with compact dimensions
included or in the dimensional reduction, so that $\text{Area}[\sigma]/(G_NM)$ is
invariant under dimensional reduction, with the truth value of the PI being the
same with or without compact dimensions.}.

\textit{Constructing Initial Data.} All the quantities appearing in the Penrose
inequality can be located on a single timeslice, so we can test the Penrose
inequality with initial datasets rather than full spacetimes.
Let us now describe how we construct initial data.  A spacelike initial
dataset for the Einstein-Klein-Gordon system on a manifold $\Sigma$ at a moment of time symmetry 
consists of a Riemannian metric $\gamma_{ab}$ and a scalar profile $\phi$ on
$\Sigma$ that together satisfy the Einstein constraint equations.
The extrinsic curvature $K_{ab}$ and time-derivative of $\phi$ on $\Sigma$ are
both vanishing. In this case, the full constraint equations reduce to
\begin{equation}
\begin{aligned}
    \mathcal{R} + \frac{ d(d-1) }{ L^2 } = |\nabla \phi|^2 + 2V(\phi),
\end{aligned}
\end{equation}
where $\mathcal{R}$ is the Ricci scalar of $\gamma_{ab}$. 

Next, we want the initial data to have finite mass
and evolve to an AAdS spacetime, which constrains $\phi$ to 
fall off sufficiently fast. Furthermore, we demand $\sigma$
to be outermost minimal, so that we can test Eq.~\eqref{eq:AdSPI}. 
Note that $K_{ab}=0$ implies that $\sigma$ is 
extremal, so we need not impose the condition that $\sigma$ can be perturbed
inwards to a trapped surface.

To make the procedure explicit, we pick our coordinate system on $\Sigma$ to be
\begin{equation}
\begin{aligned}
    \dd s^2 &= \frac{\dd r^2}{1 + \frac{ r^2 }{ L^2 } - \frac{
        \omega(r) }{ r^{d-2}}} + r^2 \dd \Omega^2, \quad r \in [r_0,
    \infty), \label{eq:cancoords}
\end{aligned}
\end{equation}
where $\omega(r)$ is a real function 
and $\dd \Omega^2$ the metric of a round unit $(d-1)$--sphere. 
The marginally trapped surface $\sigma$ is the sphere at $r=r_0 > 0$, and since
we are considering a spacelike manifold, we need $\omega(r)\leq
r^{d-2}+r^{d}L^{-2}$.
As discussed in
\cite{EngFol21a}, the above coordinates break down only at locally stationary
spheres, where the former inequality becomes an equality.
Since we want $\sigma$ to be outermost minimal, one coordinate system of the form \eqref{eq:cancoords} must be enough to cover $\Sigma$. 
In these coordinates, for a general choice of scalar profile $\phi(r)$, the solution to the constraint reads (see for example \cite{HerHor03, EngFol21a})
\begin{align}\label{eq:scalarSols}
        \omega(r) &= e^{-h(r)}\left[ \omega(r_0) + \int_{r_0}^{r}\dd \rho \frac{e^{h(\rho)} \rho^{d-1} 
        \chi(\rho)}{d-1} \right], \\ 
        h(r) &=\int_{r_0}^{r} \dd \rho \frac{\rho
        \phi'(\rho)^2}{d-1},\ 
        \chi(r) =  \left(1 +  \frac{ r^2 }{ L^2 } \right)\phi'(r)^2 +
        2 V\left(\phi\right). \nonumber
\end{align}

To construct particular initial datasets, we must provide the profile
$\phi(r)$ on $[r_0, \infty)$, together with value for $r_0$. 
The
constant $\omega(r_0)$ is
fixed by the condition of $\sigma$ being marginally trapped, giving that
$\omega(r_0) = r_0^{d-2} + r_0^{d}L^{-2}$.
Finally, we can complete our
initial dataset by gluing a second copy
of the initial dataset to itself along $\sigma$ \footnote{
Gluing $\Sigma$ to an identical copy $\Sigma'$ leads to a kink in
$\phi(r)$ at $\sigma$, but we can smooth out this kink in an arbitrarily small
neighbourhood $U=(r_0, r_0+\mathcal{\epsilon}) \subset \Sigma'$ without
altering the initial data on $\Sigma$. This might produce a large $\phi''(r)$ in
$U$, but since $\phi''(r)$ does not appear in the constraint equations the
solution to the constraints on $\Sigma'$ still exists for sufficiently small
$\epsilon$.
}
(possible since $\sigma$ is extremal -- see \cite{EngWal17b, EngWal18} for
details).

Let us now choose concrete scalar profiles. 
Since we are looking for
counterexamples to the PI rather than a proof, we are free to
consider special initial data.
We consider two types of profiles, either
\begin{equation}\label{eq:deltaProfile}
\begin{aligned}
    \phi(r) &= \sum_{k=0}^{3}\text{sign}(\eta_k) \left(\frac{|\eta_{k}|}{r}\right)^{{\Delta+2k}},
\end{aligned}
\end{equation}  
or
\begin{equation}\label{eq:logProfile}
\begin{aligned}
    \phi(r) = \begin{cases}
    \mu \log(r/R_0) & r_0 \leq r \leq R_0 \\
    0 & R_0 \leq r
\end{cases},
\end{aligned}
\end{equation}
for general constants $\{\eta_{k}\}$ and $\{\mu, R_0\}$ parametrizing the
initial data.
After picking numerical values of $r_0$ and either $\{\eta_{k}\}$ or $\{ \mu,
R_0 \}$, we can compute the integrals \eqref{eq:scalarSols} numerically, and we can obtain the mass as $16\pi G_N M
= (d-1)\text{Vol}[S^{d-1}]\omega(\infty)$.
The only remaining thing to check is that $\omega(r)$ never exceeds
$r^{d-2} + r^{d}L^{-2}$ for $r>r_0$. As long as this is true, $\sigma$
satisfies the technical conditions required for the holographic
derivation of Eq.~\eqref{eq:AdSPI}.

Why do we choose the profiles \eqref{eq:deltaProfile} and \eqref{eq:logProfile}?
By trying to minimize the mass while holding $\text{Area}[\sigma]\propto r_0^{d-2}$
fixed, we are
maximizing the chance of violating the PI, since smaller $M$
means smaller $\text{A}_{\rm BH}(M)$. To achieve a small mass, we want large regions of nonzero scalar field
in order to accumulate negative energy through the potential, while minimizing
the positive gradient contribution from $\chi(r)$. Thus, we want a scalar that
falls off slowly and without unnecessary non-monotonic behavior.
Furthermore, due to the factor $\exp[-(d-1)^{-1}\int^{\infty}_{r}\dd \rho \rho
\phi'(\rho)^2 ]$ in the integrand of Eq.~\eqref{eq:scalarSols} when computing
$\omega(\infty)\propto M$, it is the
behavior of $\phi$ at large $r$ that matters (or the largest values of $r$ where
$\phi$ has
support). Contributions to the mass from smaller $r$ are exponentially suppressed.
Now, a logarithmic profile has a slow monotonic falloff, but it requires compact support in order to
have the requisite asymptotics. 
The profile \eqref{eq:deltaProfile} has the slowest possible falloff compatible
with non-compact support and standard Dirichlet boundary conditions.  

We now generate a particular dataset by first drawing $r_0$ with a
uniform distribution from the range $(10^{-2} L, 20L)$, allowing both
small and large black holes. For the profile \eqref{eq:deltaProfile},
we draw the coefficients $\eta_{k}$ from the range $(-3r_0, 3r_0)$,
again with a uniform distribution. For the profile
\eqref{eq:logProfile} we draw $\mu \in (0, 10)$ and $R_0 - r_0 \in (0,
100L)$. The parameter ranges are chosen partly through trial and error -- if
we increase the parameter ranges for $\eta_{k}$ or $\mu$, we mostly
produce invalid datasets where $\omega(r)\geq r^{d-2}+r^{d}L^2$ at some
finite $r>r_0$. This is not surprising, since if $\phi$ gets a large 
amplitude, $\omega'(r)$ becomes large as well, causing 
$\omega(r)$ to overshoot $r^{d-2}+r^{d}L^{-2}$ near $r_0$ \footnote{We can check
that $\omega'(r_0)\leq d r_0^{d-1} L^{-2}
+(d-2)r_0^{d-3}$ is required to avoid overshooting $r^{d-2}+r^{d}L^{-2}$ near $r=r_0$,
which gives that $\phi'(r_0)$ should be $\mathcal{O}(r_0^{-1})$,
justifying the parametric dependence on $r_0$ chosen for $\mu$ and $\eta_{k}$.}.
Either way, the extent that our
sampling of the space of profiles $\phi(r)$ is suboptimal corresponds
to how much our exclusion plots below can be improved in the future.

\textit{Coupling Exclusion Plots.}
Let us first study $d=3$ and the potential
\begin{equation}\label{eq:exampleTheory}
\begin{aligned}
    V(\phi) = -\frac{ 9 }{ 16 }\phi^2 + 9\phi^3 + 11 \phi^4,
\end{aligned}
\end{equation}
which has $m^2 = \frac{ 1 }{ 2 }m_{\rm BF}^2$. This theory does not have a superpotential,
since solving \eqref{eq:superPot} gives that a real $W(\phi)$ can only
exist on a finite interval. However, we find no negative mass solutions
after generating $10^5$ initial datasets. 
Nevertheless, this theory violates the PI. For example, the profile
\begin{equation}\label{eq:exampledata}
\begin{aligned}
    \phi(r) =  \left(\frac{ 3.8 }{ r } \right)^{\Delta}\left[-1 + \left(\frac{ 3.2 }{ r } \right)^{2}
    - \left(\frac{ 2.4 }{ r } \right)^{4}\right],
\end{aligned}
\end{equation}
with $r_0 = 2.5L$  yields 
\begin{equation}
\begin{aligned}
    A[\sigma]/A_{\rm AdS-Schwarzschild}(M) \approx 1.2, \quad G_N M \approx 7 L.
\end{aligned}
\end{equation}
As shown by Penrose's original
argument \cite{Pen73},  the dataset \eqref{eq:exampledata} cannot settle down to
a stationary black hole, so it will either
collapse to a naked singularity, or we will have a Coleman-DeLuccia type decay
\cite{ColDel80} \footnote{We thank Juan
Maldacena for suggesting this possibility.},
where the conformal boundary terminates in finite time, 
and where the event horizon grows to infinite area. 

\begin{figure}
\centering
\includegraphics[width=0.45\textwidth]{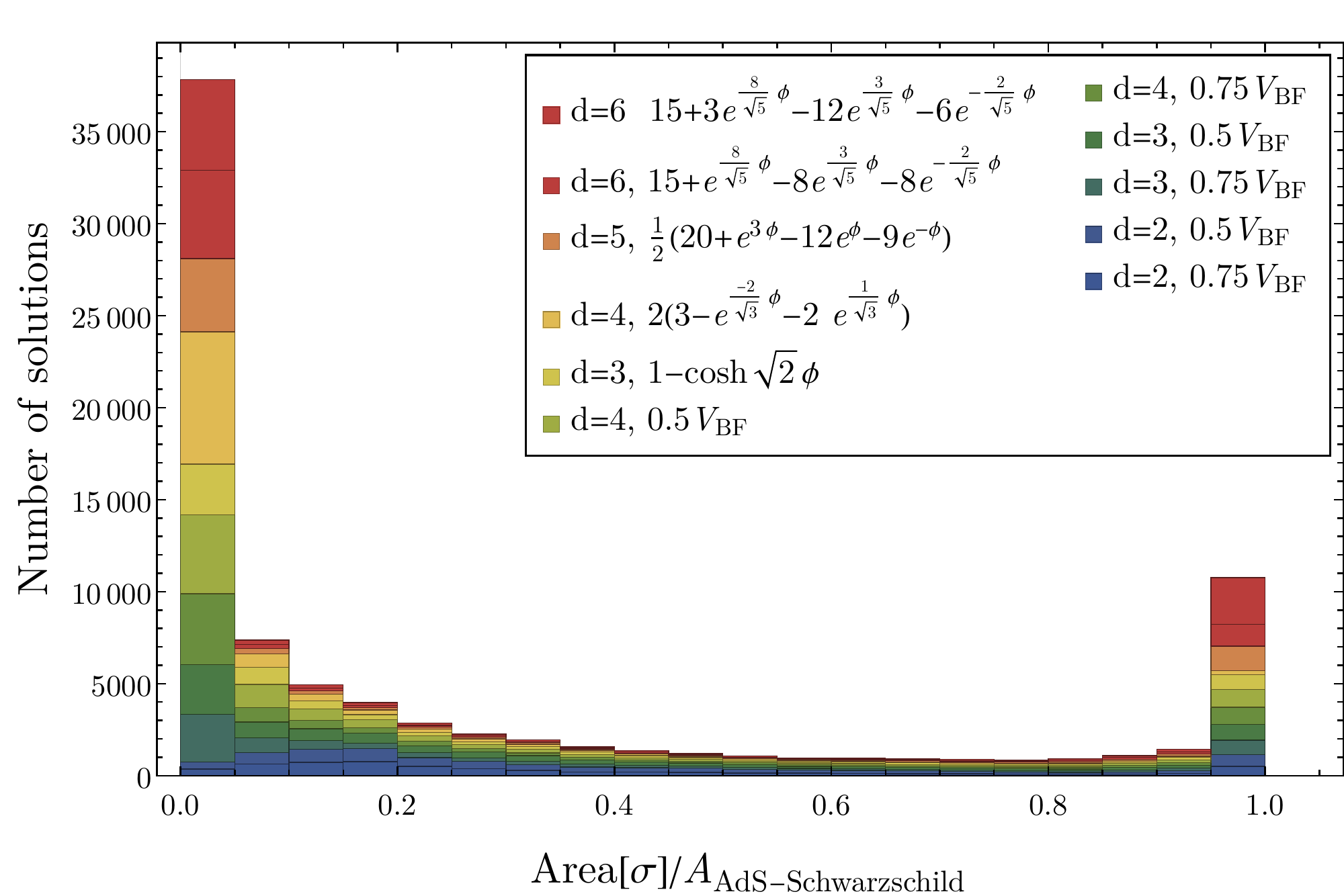}
    \caption{Plot of computed area ratios for various scalar potentials, with an
    ensemble of $10^4$ initial data sets for each potential. $V_{\rm BF}\equiv
    \frac{ 1 }{ 2 }m_{\rm BF}^2 \phi^2$. For the interacting theories,
    the $d=6$ and $d=3$ potentials come from $S^4$ \cite{LuPop99b} and $S^7$ \cite{CveDuf99}
    reduction of $D=11$ SUGRA. The
    $d=5$ potential comes from $S^4$ reduction of massive Type IIA SUGRA \cite{CveLu99}, and the
    $d=4$ potential from $S^5$ reduction of Type IIB SUGRA \cite{LuPop99}.}
\label{fig:histograms}
\end{figure}

Let us now repeat the analysis for multiple potentials. In Fig.~\ref{fig:histograms} we show
histograms of computed ratios $\text{Area}[\sigma]/A_{\rm AdS-Schwarzschild}(M)$
in a large ensemble of initial datasets with
potentials coming either from (1) dimensional reduction of SUGRA
theories appearing in string theory and AdS/CFT, such as $D=11$
\cite{LuPop99b, CveDuf99}, Type IIB \cite{LuPop99}, or massive Type IIA
\cite{CveLu99} SUGRA, or (2) corresponding to a free tachyonic scalars with $m^2
> m_{\rm BF}^2$. In the case of SUGRA, since we use scalar theories arising from consistent truncations, our
initial datasets provide valid initial datasets in the various SUGRA theories,
both in the dimensional reduction and with compact dimensions included (using
the embeddings in \cite{LuPop99b, CveDuf99, LuPop99, CveLu99}). The specific potentials are shown in the legend of
Fig.~\ref{fig:histograms}.
We see that the \PI holds for all our initial datasets. 
This does not amount to a proof that the \PI
holds, but it provides evidence, since for other potentials we will easily be able to
produce violations while sampling from the same space of scalar profiles. This
is an important consistency check on our proposal, since if the PI was violated
for theories known to have a CFT dual, it presumably cannot serve as a constraint on low
energy theories that can arise as the low energy limit of quantum gravity (a so-called swampland
condition \cite{Vaf05,ArkMot06,OogVaf06}) \footnote{Note
that for the $d=4$ potential, we only consider the logarithmic
profile, since the potential saturates the BF bound, so the mass formula requires
modification for scalar profiles with non-compact support (see for example
\cite{HerHor04}).}.

\begin{figure}
\centering
\includegraphics[width=0.45\textwidth]{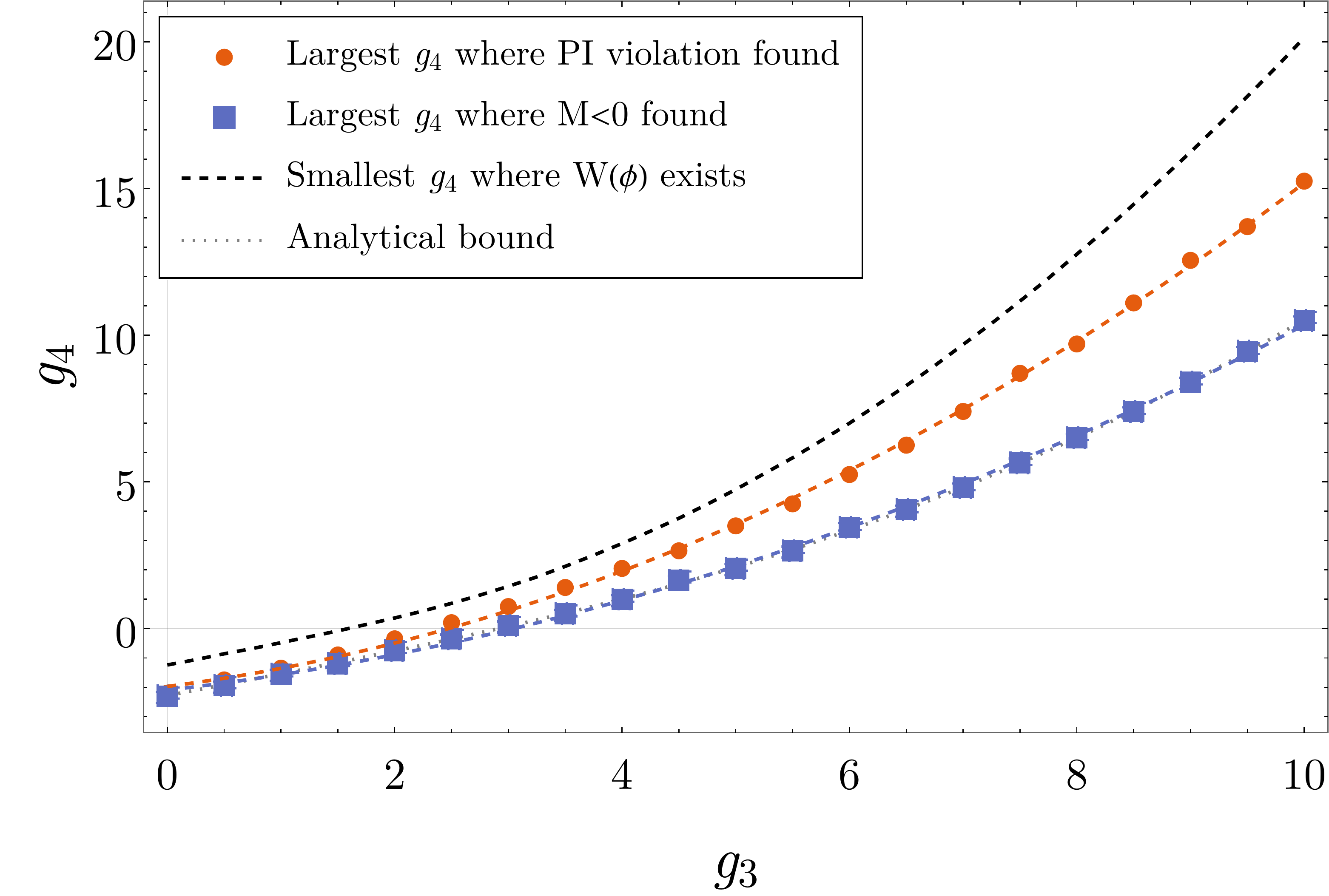}
    \caption{
    Exclusion plot on couplings for the potential $V(\phi) = \frac{ 1 }{ 4 }m_{\rm BF}^2 \phi^2 + g_3
    \phi^3 + g_4 \phi^4$ in $d+1=4$. Couplings below the circular markers are ruled
    out by the PI, while
    couplings below the squares are ruled out by positive mass. 
    Blue and orange lines are quadratic fits, and 
    couplings above the black dashed line give potentials which has
    superpotentials. 
    Above the blue and orange markers, we have found no
    violations after the construction of $10^5$ initial data sets using our
    sampling procedure.
    The dotted gray line, here coinciding with the blue, shows the exclusion boundary
    from the analytical condition \eqref{eq:poscond}. 
    }
\label{fig:exclusiong3g4}
\end{figure}

\begin{figure}
\centering
\includegraphics[width=0.45\textwidth]{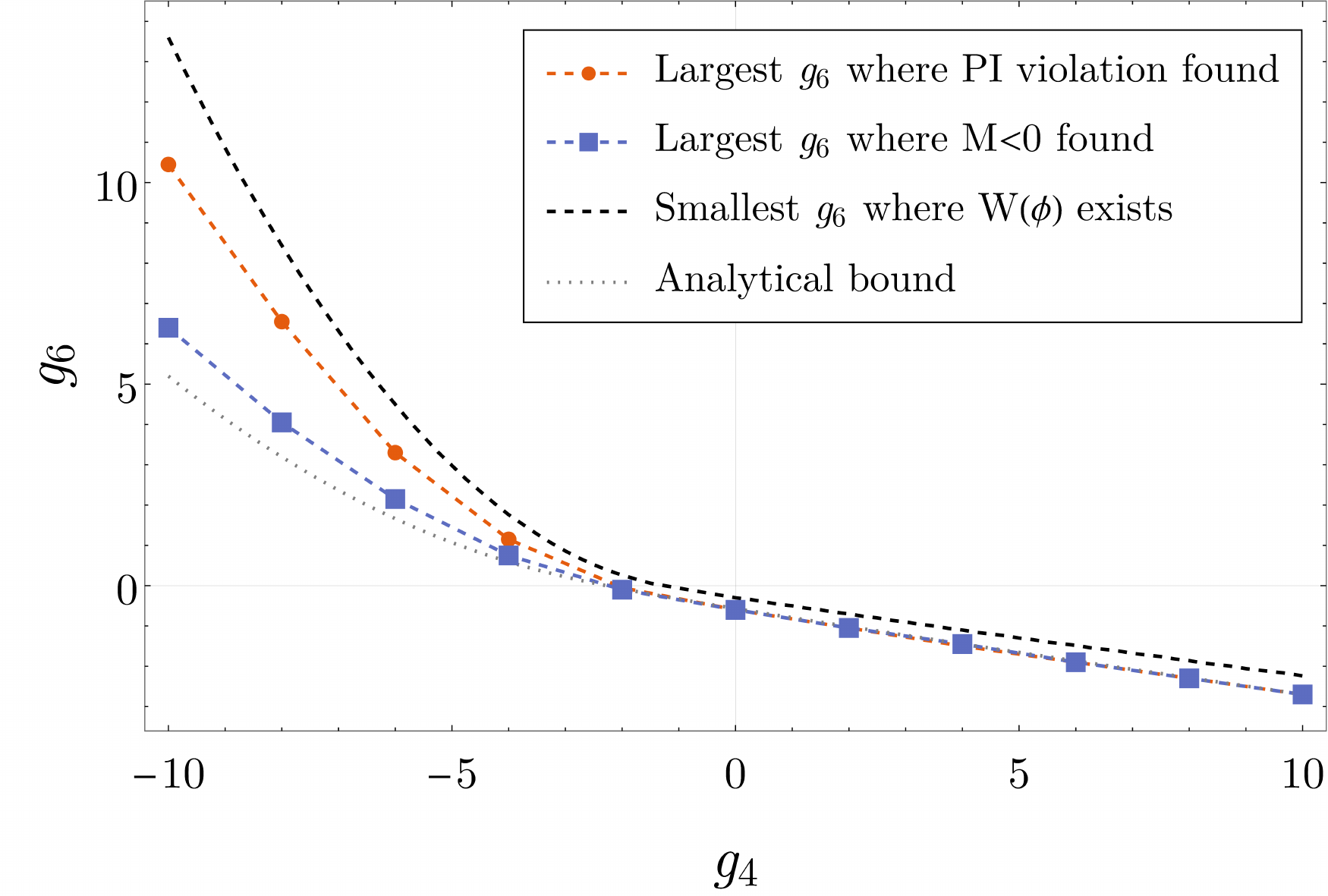}
    \caption{
    Exclusion plot on couplings for the potential $V(\phi) = \frac{ 1 }{ 4 }m_{\rm BF}^2 \phi^2 + g_4
    \phi^4 + g_6 \phi^6$ in $d+1=4$. The same description as in Fig.~\ref{fig:exclusiong3g4}
    applies, except the orange and blue lines are interpolations rather than
    fits.
    }
\label{fig:exclusiong4g6}
\end{figure}

Consider now $d=3$ and a potential with $m^2 = \frac{ 1 }{ 2
}m_{\rm BF}^2$, and with varying cubic and quartic couplings $g_3, g_4$ (see
caption of Fig.~\ref{fig:exclusiong3g4}). Take $g_3\geq
0$ without loss of generality. For a given value of $g_3$, we can gradually lower
$g_4$ until we find a dataset violating the PI or the PMT. In Fig.~\ref{fig:exclusiong3g4} we plot the
highest value for $g_4$ for which we are able to find at least one
violating dataset. Furthermore,
we plot the region in $(g_3, g_4)$ space in which a superpotential
exists. The region of
coupling space below the orange (blue) markers is ruled out by the \PI (PMT).
For $g_3>0$, the PI is a stronger condition than the
PMT -- at least in the space of initial data we are sampling. For reasons we do
not understand, at $g_3=0$ where $\mathbb{Z}_2$ symmetry is restored, 
the PI and PMT are violated at the same time. However, $\mathbb{Z}_2$ symmetry
does not appear to always guarantee coincidence, as shown in
Fig.~\ref{fig:exclusiong4g6}. Nevertheless, for $d=3$ and a potential $V=\frac{ 1 }{ 2
}m^2\phi^2 + g_4 \phi^4$, we find that the PI and PMT exclusion lines do
coincide as we vary $(m^2, g_4)$, and furthermore that exclusion line is well
described by the analytical condition given below. 

Note that there are no immediately obvious changes in the potential as we cross
the line into territory where we violate the PI. No new extrema develop. 

\textit{Analytic bounds on couplings.}
So far we have given numerical bounds on couplings, through violation
of the PI. We can also give analytical
bounds, although they are somewhat weaker, and rely on violation of the PMT
(implying PI violation). Consider the scalar profile
\eqref{eq:logProfile}, and a potential $V=\sum_{n=2}^{\infty}g_n \phi^{n}$. It
is in fact possible to solve the integrals \eqref{eq:scalarSols} analytically in
terms of gamma functions, and while the solution is somewhat involved, the
leading part of $\omega(\infty)$ in the limit $R_0 \rightarrow \infty$ is
simple, yielding, up to $\mathcal{O}(R_0^{-2})$ corrections,
\begin{align}
    \frac{ \lambda \omega(\infty) }{ R_0^{d} } &= \frac{ \mu^2 }{ L^2 }+ 2 \sum_{n=2}^{\infty} n! g_n
    \left[\frac{ (1-d)\mu }{ d(d-1) + \mu^2 }\right]^{n}, \label{eq:poscond}
\end{align}
where $\lambda \equiv  d(d-1) + \mu^2$, and with the dependence on $r_0$
contained in the $\mathcal{O}(R_0^{-2})$ terms.
A sufficient condition for violation of the PMT and PI is for the RHS of
\eqref{eq:poscond} to be negative for some $\mu \in \mathbb{R}$. Thus, any
theory where pure AdS is nonperturbatively stable must have a positive RHS of
\eqref{eq:poscond} for all $\mu$. We included the
exclusion line obtained from Eq.~\eqref{eq:poscond} in
Figs.~\ref{fig:exclusiong3g4} and \ref{fig:exclusiong4g6}.

\textit{Discussion.}
There is by now a robust trend of proposing constraints on gravity
theories in order for black holes to be well behaved semiclassically \cite{ArkMot06, BanSei10}, and for these constraints to later be proven 
in holography \cite{HarOog18, HarOog18b, EngFol20}. While the PI can be derived
in holography, we have shown that it is generally false in GR, and argued that it serves as
a new swampland \cite{Vaf05,ArkMot06,OogVaf06} condition. As an example, we showed that it can be used to
constrain scalar potentials for theories in AdS. 
If holography makes sense in asymptotically flat space, 
 it is possible that the same logic can be applied there.

\begin{acknowledgments}
\textit{Acknowledgments.}
    It is a pleasure to thank Netta Engelhardt and Gary Horowitz for comments on an earlier draft of this
    manuscript, and  Aditya Dhumuntarao, Netta Engelhardt, Gary Horowitz, Veronika Hubeny,
    Marcus Khuri, Juan Maldacena, Don Marolf, and Cumrun Vafa for discussions. My research is
    supported in part by the John Templeton Foundation via the 
    Black Hole Initiative, NSF grant no. PHY-2011905, and an Aker
    Scholarship. This research was also supported in part by the
    Heising-Simons Foundation, the Simons Foundation, and NSF grant no. PHY-1748958.
\end{acknowledgments}

\bibliography{all}
\end{document}